# Strain induced orbital dynamics across the Metal Insulator transition in thin VO$_2$/TiO$_2$(001) films

A. D'Elia[a,b,*], S.J. Rezvani[c,b], A. Cossaro[b], M. Stredansky[a], C. Grazioli[b], B. W. Li[d], C.W. Zou[d], M. Coreno[e,c], A. Marcelli[c,e,f]

a. Department of Physics, University of Trieste, Via A. Valerio 2, 34127 Trieste, Italy
b. IOM-CNR, Laboratorio TASC, Basovizza SS-14, km 163.5, 34012 Trieste, Italy;
c. Istituto Nazionale di Fisica Nucleare, Laboratori Nazionali di Frascati, 00044 Frascati, Italy;
d. National Synchrotron Radiation Laboratory, University of Science and Technology of China, Hefei 230029, P. R. China
e. ISM-CNR, Istituto Struttura della Materia, LD2 Unit, Basovizza Area Science Park, 34149 Trieste, Italy
f. Rome International Centre for Material Science Superstripes, RICMASS, Via dei Sabelli 119A, 00185 Rome, Italy

*Corresponding author: delia@iom.cnr.it

## Abstract

VO$_2$ is a strongly correlated material, which undergoes a reversible metal insulator transition (MIT) coupled to a structural phase transition upon heating (T= 67° C). Since its discovery the nature of the insulating state has long been debated and different solid-state mechanisms have been proposed to explain its nature: Mott-Hubbard correlation, Peierls distortion or a combination of both. Moreover, still now there is a lack of consensus on the interplay between the different degrees of freedom: charge, lattice, orbital and how they contribute to the MIT. In this manuscript we will investigate across the MIT the orbital evolution induced by a tensile strain applied to thin VO$_2$ films. The strained films allowed to study the interplay between orbital and lattice degrees of freedom and to clarify MIT properties.

## Introduction

The vanadium ([Ar] 3d$^3$4s$^2$) element is very reactive and can be synthesized in many mixed valence oxides with different oxidation states and stoichiometry, e.g., VO, V$_2$O$_3$, V$_3$O$_5$, VO$_2$, V$_6$O$_{13}$, V$_4$O$_7$ and V$_2$O$_5$. This class of oxides bear the seed of a strong electronic correlation and have been widely studied since the early times of X-ray spectroscopy. The electronic transport and the spectroscopic properties of V$_2$O$_3$ [1], VO$_2$ [2] as well as the complex local structure of V$_2$O$_5$ [3] have always attracted interest and still today their investigation is a hot topic in materials science, [4–6] with many potentials for technology applications [7]. Among the vanadium oxides, VO$_2$ is one of the most stimulating system. It exhibits a reversible, temperature triggered (67° C) metal insulator transition (MIT) coupled to a structural phase transition (SPT) from the high temperature tetragonal metal phase to the low temperature monoclinic insulator phase characterized also by a clear nanoscale phase separation [7–9]. Since its discovery in the late 50s [10] the nature of the VO$_2$ MIT has been object of debate within the scientific community. A transition driven by the strong electron correlation, i.e., the Mott-Hubbard transition [11, 12], or by the Peierls structural distortion [13–15] or by a cooperative Mott-Peierls mechanism are the most favoured models for VO$_2$ MIT [16]. While the structurally induced effects on the electronic properties of the materials, in particular at low dimensions and high strains are well known, [17–20] a clear correlation among lattice, orbital and electronic degrees of freedom and MIT feature is still missing.

In this structure each metal site is surrounded by slightly distorted oxygen octahedral and the crystal field splits the degenerate 3d manifold into 3 $t_{2g}$ and 2 $e^\sigma_g$ levels. The small orthorhombic distortion further splits the 3 $t_{2g}$ levels in one singly degenerate $a_{1g}$ and two $e^\pi_g$ levels. According to the Goodenough model [14] V 3d and O 2p orbitals hybridize forming bonds of σ and π symmetry. Their unoccupied levels are identified as π*($e^\pi_g$ character) and σ* ($e^\sigma_g$ character). The $a_{1g}$ orbital is populated by unpaired 3d electrons and is called $d_\parallel$. The dimerization of vanadium atoms in the insulating phase splits the $d_\parallel$ originating empty $d^*_\parallel$ (with $t_{2g}$ character). In the metallic phase the $d^*_\parallel$ is oriented along the $c_r$ axis and along the V-V dimer in the insulating phase. This bond is strictly related with the unidimensional V-V dimer chain formation in the monoclinic insulating phase [21], while the π* has an isotropic behaviour within the lattice [22]. Across the MIT the π* and $d^*_\parallel$ collapse to the Fermi level (FL) upon being both populated. This mechanism closes the band gap. Moreover, since π*, $d^*_\parallel$ and σ* have mostly a 3d character, changes on the electronic structure can be followed by monitoring the absorption spectroscopy intensity at the V L edges (V 2p -> 3d)[8]. Here we present the results obtained on three different single crystalline strained films of VO$_2$/TiO$_2$(001) with thickness 8, 16 and 32 nm, probing simultaneously the structural and orbital contribution to the MIT using the XANES spectroscopy.

## Methods

Films of VO$_2$ having a thickness of 8, 16 and 32 nm were deposited on a clean substrate of TiO$_2$ (001) by the RF-plasma assisted oxide-MBE instrument working with a base pressure better than 4x10$^{-9}$ mbar. At a constant growth rate of 0.1 Å/s, the thickness was controlled by adjusting the deposition time in a range from several unit cells to tens of nanometers. During the deposition process, the substrate has been kept at the temperature of 550 °C. The interfacial cross-section has been investigated with the high-resolution scanning transmission electron microscope (STEM). High angle annular dark field (HAADF) STEM images were taken on a JEM ARM200F with a probe aberration corrector, while the diffraction pattern was acquired on a JEM 2100 TEM. The complete details of the epitaxial film preparation are reported elsewhere [23, 24].

The XANES experiments have been performed at the ANCHOR end-station of the ALOISA beamline [25] at Elettra synchrotron radiation facility. Electrons were collected at normal emission by a PSP Vacuum 120 mm electron analyser with 2D delay line detector. The photon beam was linearly polarized in the scattering plane and impinging the sample at the magic angle (35°). Measurements were performed at constant pass energy ($E_p$=20 eV).

## Results

To discern the spectral changes observed in the XANES spectra of these films it is necessary to understand the strain-induced modification of the VO$_2$ crystal structure. The TiO$_2$ substrate has the tetragonal (rutile) lattice structure as the metallic VO$_2$. Most of the considerations in the next are referred to the rutile phase of VO$_2$ except where specified otherwise. Since the TiO$_2$ substrate is oriented along the (001) surface, the lattice mismatch will affect the *a* and *b* structural parameters of the vanadium oxide films. The in plane lattice mismatch can be calculated as:

$$M = \frac{a_s - a_f}{a_f} * 100\% \qquad (1)$$

where $a_s$ and $a_f$ are the lattice parameters of the substrate and of the sample, respectively. Between rutile TiO$_2$ (a=b=4.58 Å) and bulk VO$_2$ (a=b=4.55 Å) the lattice mismatch M is 0.66%. Then to match the substrate lattice during the early stage of the epitaxial growth, a VO$_2$ film will undergoes to a tensile strain, which results in the increase of both $a_r$ and $b_r$ and the consequent elastic compression of $c_r$ [22, 23, 26]. Increasing the thickness of the VO$_2$ film, the distortions induced by the lattice of substrate fades and the lattice constants relax to the bulk VO$_2$ values. For the sample analyzed in this work, the critical thickness for which the sample can be considered bulk-like is ~25 nm [23].

In the tetragonal unit cell, the vanadium atoms occupy the positions (0,0,0) and (½, ½, ½). Each vanadium atom is surrounded by oxygen octahedron with two different V-O bond lengths. Metal and oxygen atoms separated by the apical distance, share the same $z$ value along $c_r$. The equatorial distance that separates the vanadium atom and the four neighboring oxygen atoms is $z = z_{metal} \pm ½$ (see Figure 1). The two apical oxygen atoms are located at ± (u, u, 0) while the four equatorial oxygen atoms are in the positions ±(±(u-½), ∓(u-½); ½) where u=0.3001 at 360 K [13, 27] although V-O bond lengths are influenced by strain-induced modifications in the lattice parameters. Looking at the apical distance this is described by:

$$J_{Apical} = J_A = \sqrt{u^2 a_r^2 + u^2 b_r^2} = \sqrt{2u^2 a_r^2} = \sqrt{2} u a_r \quad (2)$$

and the apical V-O bond increases linearly with the $a_r$ length. On the other hand, the equatorial bond length includes the three lattice parameters:

$$J_{Equatorial} = J_E = \sqrt{(u - \tfrac{1}{2})^2 a_r^2 + (u - \tfrac{1}{2})^2 b_r^2 + \tfrac{1}{4} c_r^2} = \sqrt{2(u - \tfrac{1}{2})^2 a_r^2 + \tfrac{1}{4} c_r^2} \quad (3)$$

Combining equations 2 and 3 we may recognize that $J_A$ increases with the strain, while $J_E$ is almost independent since the increase in $a_r$ is compensated by the decrease of $c_r$ [22]. Moreover, increasing the apical V-O distance the superposition between oxygen and vanadium orbitals decreases and as a consequence the 3d-2p hybridization.

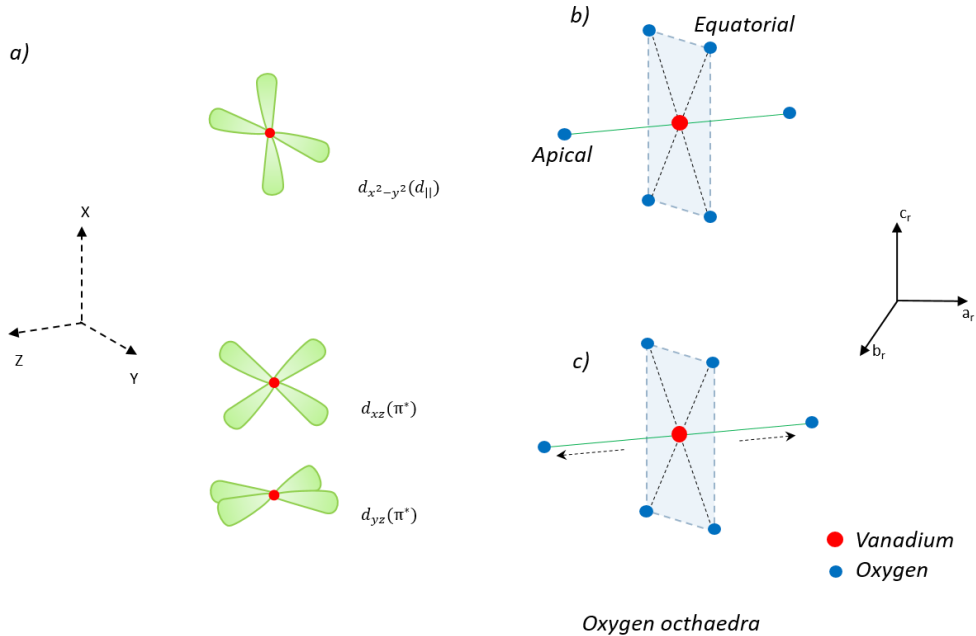

Figure 1 a) Representation of the π* and $d_{\parallel}^*$ orbitals in the octahedral frame of reference XYZ. b) Oxygen octahedron surrounding V atom in the bulk case. c) Schematic representation of the strain effect on the oxygen octahedron. The mismatch between TiO$_2$ ($a_r=b_r$=4.58 Å) and VO$_2$ ($a_r=b_r$=4.55 Å) increases the $a_r$ and $b_r$ lattice parameters in the epitaxial film while decreasing $c_r$. This results also in the increase of the apical V-O bond length. The reference frame of the octahedron is rotated by 45° respect to the tetragonal unit cell ($a_r b_r c_r$).

Actually, the π* orbital, which points toward other vanadium atoms and it is oriented toward the oxygen corners of the octahedron, is the most affected bond [22]. Moreover, the decrease of the V-O hybridization reduces the bonding-antibonding energy separation, hence the energy of π* orbital lowers. The $d^*_{||}$ experiences the opposite situation. The decrease of $c_r$ increases the overlap among orbitals within the unidimensional V-V chains, shifting to high energy the $d^*_{||}$ orbital. Indeed, the strain increases the π*-$d^*_{||}$ splitting reversing the orbital population at the FL. In the insulating phase, the V-O hybridization is stronger and, as a consequence
the π* and $d^*_{||}$ orbital are further separated, with the second appearing at higher energy respect to the former.

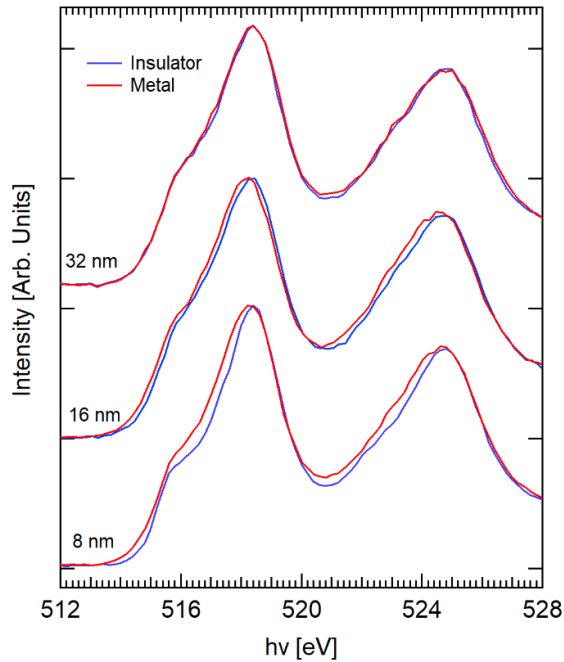

*Figure 2 Comparison among XANES spectra (Auger V $L_3M_{23}M_{45}$, 464 eV) in the hv range (512-528 eV) of the insulating (blue line 30° C) and metallic (red line 90 °C) phase of different $VO_2$ samples. From top to bottom: strained samples of 32, 16 and 8 nm. The spectra are normalized to the incident photon flux and to the $L_3$ maximum.*

The V L edges XANES spectra for the metallic and insulating phase are compared in Figure 2.

The interpretation of the L edge shape is not straightforward. However, the $L_3$ edge exhibits more defined features with respect to the $L_2$.one. In general the $L_3$ and $L_2$ spectra differs because of the multiplet effects in the final state [28]. In this particular case these are negligible, whereas the presence of a strong Coster–Kronig decay (V $L_2L_3M_{4,5}$ ) [29] severely reduce the life-time of the excited state $2p^1_{1\backslash2}$ $3d^{n+1}$, thus broadening the XANES line shape and making hard to recognize different adjacent features. As a consequence, we will focus our attention to the $L_3$ edge.

The shape of the $L_3$ edge is not commensurate with the shape of the O K edge spectrum available in literature [30, 31] despite they should exhibit the same features as a consequence of the V3d- O2p hybridization. This can be understood taking into account other effects: in the $L_3$ and $L_2$ XANES there is a transfer of spectral weight away from threshold ($L_3$ maximum intensity 518.4 eV), the apparent reduction of the spin-orbit splitting (6.4 eV from XANES spectra, 7.3 eV from XPS [32]) and the deviation from the statistic intensity ratio $I(L_3)/I(L_2)=2$. All features can be explained considering the strong interaction between the 2p core hole (**2p**) and the 3d electrons in the final state. In vanadium oxides **2p**-3d interaction is of the same order of magnitude of the spin-orbit splitting, with a severe redistribution of the spectral weight of the entire spectrum [31, 33]. The main shape changes occur in the spectral region 514-518.5 eV where are present the π*, $d^*_{||}$ and σ* orbitals [21, 22, 34]. Precisely, π*and $d^*_{||}$ are not distinguishable, but are located in the range 514-516.5 while σ* is centered at 518.4 eV [21]. However, in the insulating phase when the strain increases, the π*- $d^*_{||}$ features become more evident because of the increasing π*-$d^*_{||}$ splitting. To highlight the changes across the MIT, the difference spectra calculated using the Eq. 4 are showed in Figure 3.

$$I_{ins} - I_{met} \propto uDOS_{ins} - uDOS_{met} \quad (4)$$

Two main contributions can be identified at ~515 and ~517.5 eV. The high-energy contribution associated to the σ* orbital shifts toward low photon energy due to the rearrangements of vanadium atoms within the oxygen octahedron. The low energy contribution can be assigned to the π*-$d^*_{||}$ rearrangement going from the insulating to the metallic phase. Actually, the spectral difference is dominated by the σ* signal whose low energy tail may probably affect the line shape of the π*-$d_{\parallel}^*$ feature. Nevertheless, by integrating the difference spectra in the range 512-518.4 eV, a clear trend as a function of thickness emerges. The thickness dependence of the integrated intensity is alike what reported in [22] points out that the different orbital strain dynamics are qualitatively similar in agreement with the theoretical band model reported in [23] for $VO_2$/$TiO_2$(001) strained and ultra-strained films.

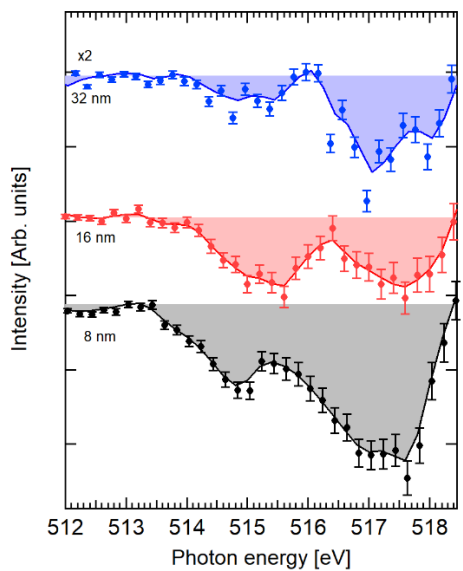

*Figure 3 Comparison among the difference of the Intensity of the XANES spectra showed in Fig. 2 in the energy range 512-518.4 eV. The dots represent the experimental points while the continuous line is the smoothed curve of the experimental points (binomial algorithm), which is used as a guide for eyes. From top to bottom: films of 32, 16 and 8 nm thickness. For sake of clarity the spectra are vertically shifted and the difference spectrum of the 32 nm film is multiplied by 2.*

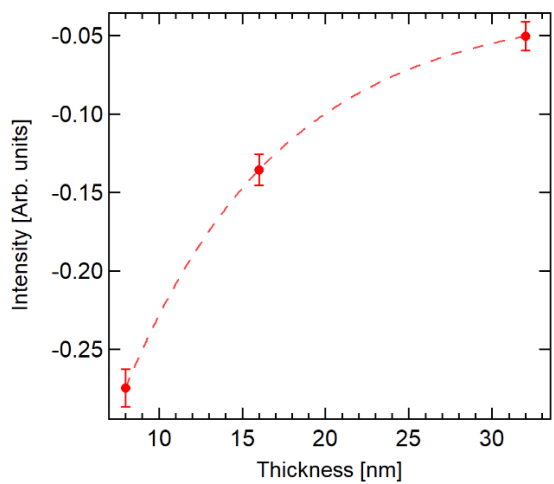

*Figure 4 The integrated differences as a function of the film thickness. The Integration of the differences has been performed using a trapezoidal algorithm and in the photon energy range 512-518.4 eV. The dashed line is a guide for the eye.*

## Conclusion

In this work we investigated three different single crystalline strained $VO_2$ films deposited over a rutile $TiO_2$(001) substrate. The epitaxially grown $VO_2$ undergoes to a tensile strain, which results in the increase of $a_r$ and $b_r$ to match the higher lattice constant of the substrate and to an elastically reduction of $c_r$. These changes in the cell unit parameters affect the oxygen octahedron, which surrounds each vanadium atoms increasing the apical V-O bond length decreasing also the V 3d-O 2p hybridization. In the metallic phase of $VO_2$ the induced strain reduces the hybridization, upshifts of the $d_{\parallel}^*$ empty orbital and downshifts of π* leading to an inversion of the $d_{\parallel}^*$/ π* occupation at the Fermi level. In the insulating phase the π* and $d_{\parallel}^*$ orbital split is augmented respect to the bulk case. In order to probe the dynamical model of the orbital states, we performed XANES measurements of the V L edges probing the Auger yield (464 eV). Across the phase transition we observe that the major changes are observed in the π*-$d_{\parallel}^*$ region of the spectra with the intensity increasing simultaneously to the strain. A numerical integration in the spectral region most perturbed by the MIT is in a qualitative agreement with the theoretical models and previous measurements, confirming the reliability of XANES measurements to probe the orbital strain dynamics across the MIT. Further theoretical and experimental investigations are necessary to improve our understanding of the interplay between orbital and lattice structure in $VO_2$ MIT.

**Conflict of Interest**: The authors declare that they have no conflict of interest.